\PassOptionsToPackage{unicode}{hyperref}
\PassOptionsToPackage{hyphens}{url}
\PassOptionsToPackage{dvipsnames,svgnames,x11names}{xcolor}
\documentclass[
  11pt,
]{article}

\usepackage{amsmath,amssymb}
\usepackage[T1]{fontenc}
\usepackage[utf8]{inputenc}
\usepackage{textcomp}
\IfFileExists{upquote.sty}{\usepackage{upquote}}{}
\IfFileExists{microtype.sty}{
  \usepackage[]{microtype}
  \UseMicrotypeSet[protrusion]{basicmath} 
}{}
\makeatletter
\@ifundefined{KOMAClassName}{
  \IfFileExists{parskip.sty}{%
    \usepackage{parskip}
  }{
    \setlength{\parindent}{0pt}
    \setlength{\parskip}{6pt plus 2pt minus 1pt}}
}{
  \KOMAoptions{parskip=half}}
\makeatother
\usepackage{xcolor}
\usepackage[margin=1in]{geometry}
\ifx\paragraph\undefined\else
  \let\oldparagraph\paragraph
  \renewcommand{\paragraph}[1]{\oldparagraph{#1}\mbox{}}
\fi
\ifx\subparagraph\undefined\else
  \let\oldsubparagraph\subparagraph
  \renewcommand{\subparagraph}[1]{\oldsubparagraph{#1}\mbox{}}
\fi

\usepackage{longtable,booktabs,array}
\usepackage{calc} 
\usepackage{etoolbox}
\makeatletter
\patchcmd\longtable{\par}{\if@noskipsec\mbox{}\fi\par}{}{}
\makeatother
\IfFileExists{footnotehyper.sty}{\usepackage{footnotehyper}}{\usepackage{footnote}}
\makesavenoteenv{longtable}
\usepackage{graphicx}
\makeatletter
\def\maxwidth{\ifdim\Gin@nat@width>\linewidth\linewidth\else\Gin@nat@width\fi}
\def\maxheight{\ifdim\Gin@nat@height>\textheight\textheight\else\Gin@nat@height\fi}
\makeatother
\setkeys{Gin}{width=\maxwidth,height=\maxheight,keepaspectratio}
\makeatletter
\def\fps@figure{htbp}
\makeatother

\makeatletter
\@ifpackageloaded{caption}{}{\usepackage{caption}}
\AtBeginDocument{%
\ifdefined\contentsname
  \renewcommand*\contentsname{Table of contents}
\else
  \newcommand\contentsname{Table of contents}
\fi
\ifdefined\listfigurename
  \renewcommand*\listfigurename{List of Figures}
\else
  \newcommand\listfigurename{List of Figures}
\fi
\ifdefined\listtablename
  \renewcommand*\listtablename{List of Tables}
\else
  \newcommand\listtablename{List of Tables}
\fi
\ifdefined\figurename
  \renewcommand*\figurename{Figure}
\else
  \newcommand\figurename{Figure}
\fi
\ifdefined\tablename
  \renewcommand*\tablename{Table}
\else
  \newcommand\tablename{Table}
\fi
}
\@ifpackageloaded{float}{}{\usepackage{float}}
\floatstyle{ruled}
\@ifundefined{c@chapter}{\newfloat{codelisting}{h}{lop}}{\newfloat{codelisting}{h}{lop}[chapter]}
\floatname{codelisting}{Listing}

\makeatother
\makeatletter
\@ifpackageloaded{caption}{}{\usepackage{caption}}
\@ifpackageloaded{subcaption}{}{\usepackage{subcaption}}
\makeatother
\makeatletter
\@ifpackageloaded{tcolorbox}{}{\usepackage[skins,breakable]{tcolorbox}}
\makeatother
\makeatletter
\@ifundefined{shadecolor}{\definecolor{shadecolor}{rgb}{.97, .97, .97}}
\makeatother
\ifLuaTeX
  \usepackage{selnolig}  
\fi
\IfFileExists{bookmark.sty}{\usepackage{bookmark}}{\usepackage{hyperref}}
\IfFileExists{xurl.sty}{\usepackage{xurl}}{} 
\hypersetup{
  pdftitle={A World in Print},
  pdfauthor={Johan Heinsen and Camilla Bøgeskov},
  colorlinks=true,
  linkcolor={blue},
  filecolor={Maroon},
  citecolor={Blue},
  urlcolor={Blue},
  pdfcreator={LaTeX via pandoc}}

\title{A World in Print}
\usepackage{etoolbox}
\makeatletter
\providecommand{\subtitle}[1]{
  \apptocmd{\@title}{\par {\large #1 \par}}{}{}
}
\makeatother
\subtitle{Introducing a Danish-Norwegian corpus of historical
newspapers}
\author{Johan Heinsen and Camilla Bøgeskov}
\date{2025-09-02}

\begin{document}
\maketitle
\ifdefined\Shaded\renewenvironment{Shaded}{\begin{tcolorbox}[sharp corners, breakable, interior hidden, borderline west={3pt}{0pt}{shadecolor}, boxrule=0pt, enhanced, frame hidden]}{\end{tcolorbox}}\fi

\hypertarget{abstract}{%
\section{\texorpdfstring{\textbf{Abstract}}{Abstract}}\label{abstract}}

This Data Descriptor introduces the dataset Enevældens Nyheder Online
(News during Absolutism Online). The Enevældens Nyheder Online (ENO)
dataset provides a reconstruction of the contents of major newspapers in
Denmark and Norway during the period of Absolutism (1660--1849). The
dataset contains approx. 474 million words, created using neural
networks designed to process digitised microfilm versions of Danish
newspapers as well as a smaller selection of Norwegian publications that
were all hitherto illegible for computers. The contributions details
this process and its results, including a way to derive standalone texts
from the editions, and the accompanying BERT-model trained on a
beta-version of the dataset.

\hypertarget{background}{%
\section{\texorpdfstring{\textbf{Background}}{Background}}\label{background}}

During the eighteenth century, newspapers became the most important
source of information to commoners in the conglomerate state of
Denmark-Norway (united until 1814). From humble beginnings in the late
seventeenth century, this cheap print media came to be an everyday
fixture in how people interpreted their worlds and engaged in society.
Whether processing political upheaval or natural disasters or navigating
the steady beats of local markets for goods and labour, the newspapers
were crucial for the distribution of knowledge. While the earliest
newspapers formed a narrow medium with a propagandistic touch, they soon
developed into being a widely distributed source of all kinds of
information relevant to many segments of society. For scholars today,
their value is indisputable as they reflect what contemporaries
considered important, and served as central venues for public
discussions.

Today, Denmark's and Norway's royal libraries each house their
respective parts of what is left of this paper horizon. A lot of it has
survived, with many series still existing in full. However, past choices
inflect on how the newspapers are available as digital text data. In
Norway, all editions have been photographed using modern equipment and
processed using standard OCR-solutions. They are available via fulltext
search in the Nettbiblioteket interface.\footnote{\url{https://www.nb.no/search}}
In Denmark, an extensive process of photographing and microfilming the
collection began already in the 1950s, and when a digital facsimile was
to be created after the turn of the century, decisions were made not to
re-photograph the pages. Therefore Denmark's Royal Library now provides
access to the material via their Mediestream platform, but only to a
digitized version of the microfilm (figure 1), which has, again, been
OCR'ed using off-the-shelf technologies.\footnote{\url{https://www2.statsbiblioteket.dk/mediestream/}}
The result is an uneven mess, to put it mildly. While the Norwegian
collection overall fares somewhat better due to the investment in new
photographs with modern equipment, both contain text data that is
typically illegible if read without reference to the original image.
Accessing the material in the Danish collection through the accompanying
API, users are provided with a key point of metadata -- a predicted word
accuracy. Generally it hovers around 50\%. A few examples of the median
pwa on titles covered in our dataset include Aarhus Stifts-Tidende:
48.5\%; Aalborg Stifts Adresseavis: 47.6\%; and Berlingske Tidende:
47\%. The Royal Library does not provide information on how this score
is calculated, but our experience in working with the data tells us that
it is fairly accurate. Despite the better condition of the Norwegian
material and the potentially numerous hits of fulltext search, one is
still left without a sense of how much the search might have missed.
Hence, as textual data for text mining or for language modelling, the
material in both collections is largely useless. This is a shame. The
poor digitisation causes the potential of this vast historical material
to remain underutilised. Elsewhere, where different digitisation
approaches of newspaper archives have resulted in more successful
transformations of newspapers into textual data, such materials have
proven to be a rich source for serial information on a wide range of
topics, as well as a testbed for new computational approaches to
historical analysis.\footnote{Ehrmann, M., Bunout, E. \& Clavert F.
  (eds.), \emph{Digitsed Newspapers - A New Eldorado for Historians?
  Reflections on Tools, Methods and Epistemology} (De Gruyter, 2023).}\footnote{Yang,
  T., Torget, A. J. \& Mihalcea, R. Topic Modeling on Historical
  Newspapers. \emph{Proceedings of the 5th ACL-HLT Workshop on Language
  Technology for Cultural Heritage, Social Sciences, and Humanities},
  96--104 (2011).}\footnote{Tantner, A. in \emph{News Networks in Early
  Modern Europe} Intelligence Offices (ed.~Tantner, A., Moxham, N. \&
  Raymond, J.) Ch. 19 (Brill, 2016).}\footnote{Resch, C., Rastinger, N.
  C. \& Kirchmair, T. From Semi-structured Text to Tangible Categories:
  Analysing and annotating death lists in 18th century newspaper issues,
  \emph{DHQ: Digital Humanities Quarterly} 17:3 (2023).} It all hinges
on the original material being rendered into somewhat accurate digital
text.

A range of factors contributes to the limited success in the
digitisation of Danish and Norwegian newspaper collections. They include
the irreversible realities of small print on cheap paper in a
deteriorated condition. Adding to this, the Danish collection is scanned
from microfilm, and both collections have deployed commercial
OCR-solutions dating back at least a handful of years. The typography is
mostly made up of fraktur print -- a typeface that has not been widely
used for a century and which traditional commercial OCR-solutions were
never optimised to handle. The relatively complex layouts, often with
thin separators between columns, add to the failure rates, as this leads
to columns not being properly separated. So reading order is also a
challenge.

\begin{figure}

{\centering \includegraphics{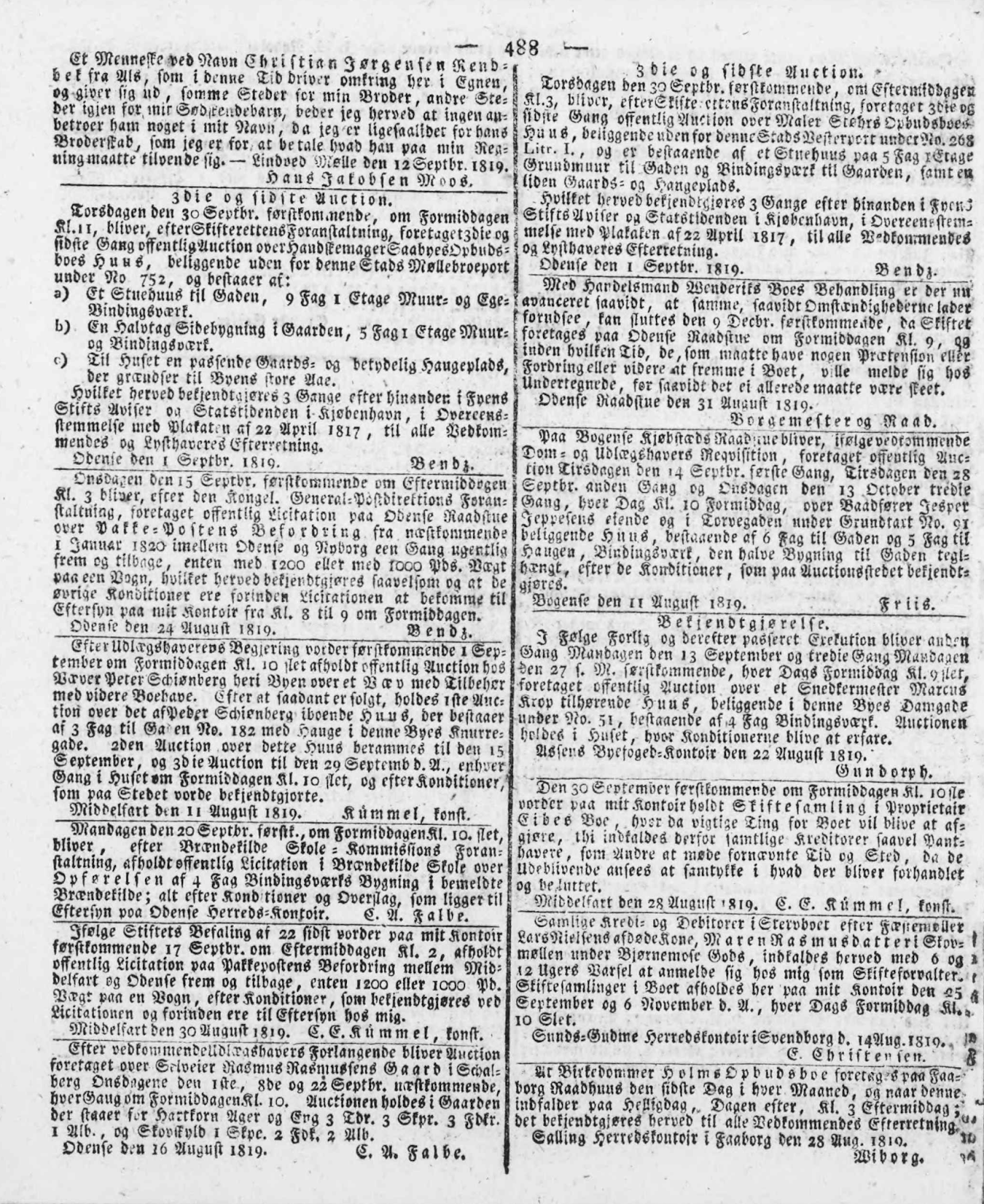}

}

\caption{A typical page with a narrow column separator. Odense
Adresse-Contoirs Efterretninger, 13 September 1819.}

\end{figure}

\hfill\break
Our dataset consists of a new processing of these older collections,
aiming to restore a more faithful digital facsimile on the basis of
existing image material. It contains 474 million words derived from
about 565,000 pages from the period. It is by far the largest fully
searchable text dataset in historical Danish dating from before 1900. In
comparison, the second largest historical corpus of a similar quality
from before 1900 is about 60 million words derived from novels 1870 to
1900.\footnote{https://huggingface.co/datasets/MiMe-MeMo/Corpus-v1.1.}
We estimate that the newspaper corpus eclipses all other Danish
digitized text data of a similar quality from this period combined. The
size as well as the diversity of the material -- reflecting multiple
concurrent political, economic and social processes -- results in the
dataset providing an altogether new foundation for computational
approaches to Danish and Norwegian history in the period.

\hypertarget{methods}{%
\section{\texorpdfstring{\textbf{Methods}}{Methods}}\label{methods}}

Having trained some of the first neural models with Transkribus to deal
with older forms of Danish handwriting, we saw an
opportunity.\footnote{\url{https://www.transkribus.org/}.}\footnote{Terras,
  M., Anzinger, B., Gooding, P., Mühlberger, G., Nockels, J., Romein, C.
  A, Stauder A. \& Stauder F. The Artifical Intelligence Cooperative:
  READ-COOP, Transkribus, and the benefits of shared community
  infrastructure for automated text recognition. Preprint at
  \url{https://doi.org/10.12688/openreseurope.18747.1} (2025).}\footnote{Prokop,
  H. \& Romein, C. A. Einsatz von künstlicher Intelligenz bei der
  automatischen Handschrifterknnung. Das Beispil Transkribus.
  proceedings from the Landesarchivtag Sachsen-Anhalt, 113-130 (2024).}
If emerging technologies could handle the high variation of handwriting,
they might also be a match for old fraktur. In early 2022, we designed a
prototype model based on about 100,000 words from eighteenth-century
pages in the Danish collection. To our surprise, its performance
exceeded expectations. On equivalent validation material, the model
performed with an error rate below 1\% on character level, translating
to a word accuracy above 95\%. That means -- in almost all cases -- a
text that is legible as is. Encouraged by these results, we started to
create a dataset of re-processed pages of the publication Københavns
Adresseavis, the main advertisement paper in Copenhagen in the period.
Running an undergraduate course in digital history at Aalborg University
in the autumn of 2023, we expanded the model capabilities by adding new
training data from other papers in collaboration with students. 10 years
of the main Aalborg newspaper from 1818 to 1827 were re-processed as a
further test. During this, the model proved effective in handling text
from different types of papers across the period. We therefore added an
even wider range of publications to the training material, and
systematically started to re-process all newspapers dating back before
1800.\footnote{\emph{Danish Newspapers 1750-1850},
  \url{https://app.transkribus.org/models/public/text/danish-newspapers-1750-1850}.}

The transcriptions are diplomatic, meaning we have transcribed each
letter exactly as is, including idiosyncratic elements of print in the
period, such as interchangeable ``i'' and ``j''. However, because our
models tend to see more of one spelling variation than another, they
occasionally perform a slight normalisation of their own. The training
material is mostly fraktur types, but the newspapers include enough
intermittent antiqua types (the type of serif types we commonly see
today) to be reasonably capable of dealing with these too. The exception
is longer series of capitalised letters where the output tends to fall
apart. Use cases that work with obituaries have to take this into
account as some papers tended to render names in obituaries in
capitalised antiqua. Similarly some ornamental types in advertisements
from the final decades of the period suffer from higher error rates than
running prose.~

To keep as much of the structural layout of the newspapers as possible,
and to reconstitute correct reading order, we also worked on finding
optimal workflows for layout segmentation. Initially, this hinged on
tailored line-recognition models -- basically creating a custom model
for every time a publisher changed anything substantial about the
layout. However with the release of the fields-model architecture on
Transkribus, we pivoted to this solution, which tended to produce better
column segmentation especially on complex layouts with narrow margins
between columns. When we found the models to struggle, we manually
checked the layout identification of each page. In this sense, layout
proved a more labour intensive challenge than the text itself.

The final dataset includes 28 newspapers of which most are represented
with complete series until a cutoff at the end of the period. Typically,
they are included in full as found in the library collections, meaning
that missing editions reflect lost items in the series themselves.
Appendixes are included when available, but generally these are much
less complete than the main publications. The included material
represents the major newspapers of the period.\footnote{Søllinge J. D.
  \& Thomsen, \emph{N. De danske aviser 1634-1989} (Odense
  Universitetsforlag, 1988)} For the period before 1800, all newspapers
available through the collections are included except for a select few
omissions in which we judged the original image material to be too
damaged for meaningful inclusion. For the period after 1800, the major
provincial papers of Denmark are included to provide wide geographic
coverage. We also included a few shorter series to represent less common
editorial viewpoints, including the somewhat radical peasant papers,
Jyllandsposten and Almuevennen, and a short-lived children's paper
running from 1779 to 1782. Two smaller series (Politivennen and
Corsaren) were digitised independently by students who had been part of
our workshops. They are included as well and are of similar quality.

Throughout, we refrain from sampling, except for Københavns Adresseavis
in the period from 1804 onwards. The reason for this exception is that
the scans of this paper from this period onwards reflect heavily
deteriorated microfilm and are ill-suited for our models. Combined with
the massive amount of pages from this particular paper and its uniform
focus on advertisements, we decided to focus our resources on papers
that would yield higher-quality text output. Therefore, we only sampled
every fifth year from then on.

The total coverage is shown in figure 2. The disproportionate amount of
data from the second half of the period is shown in figure 3, showing
the distribution of individual texts pr. year.

\begin{figure}

{\centering \includegraphics{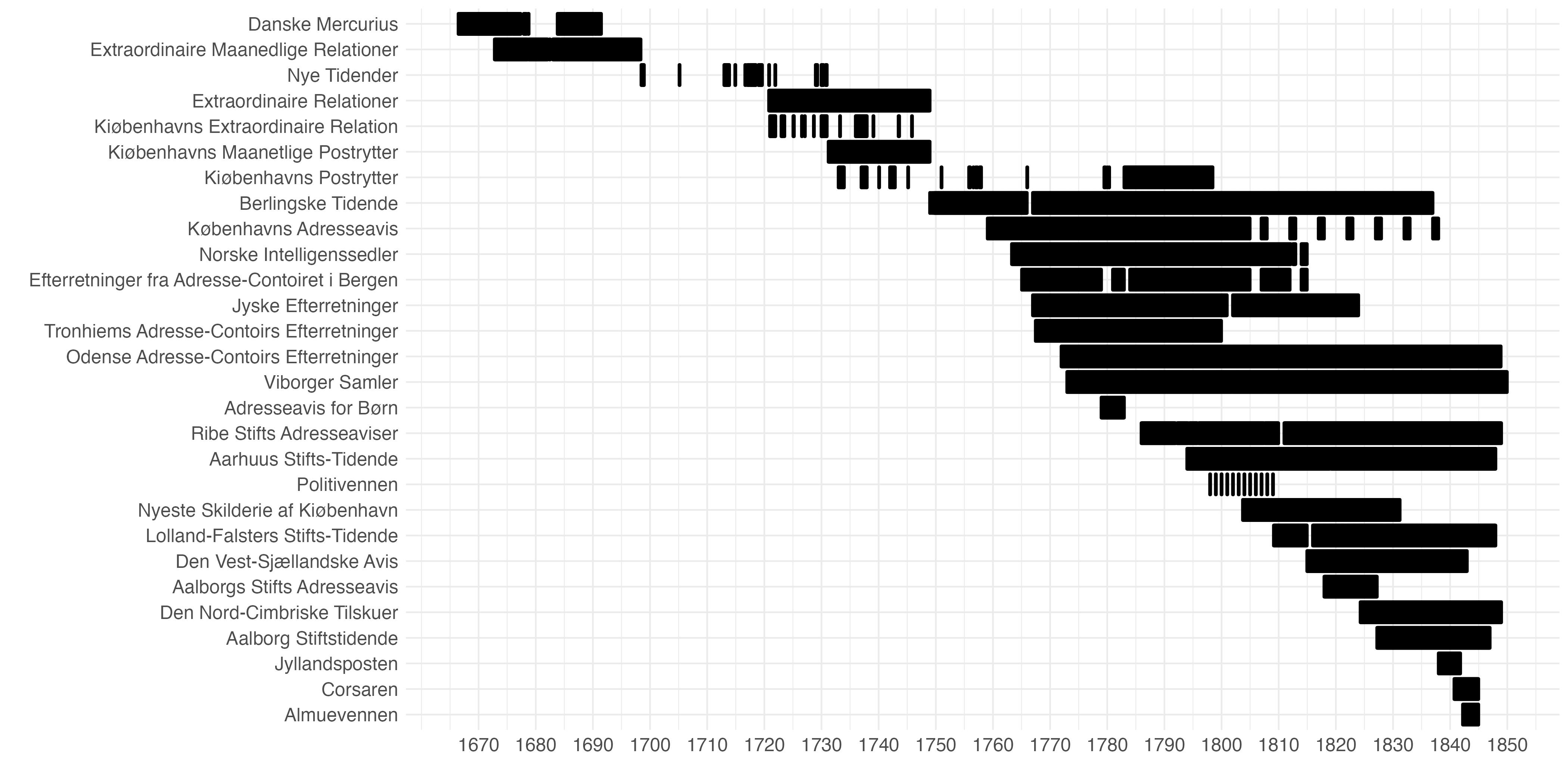}

}

\caption{Coverage over time.}

\end{figure}

\begin{figure}

{\centering \includegraphics{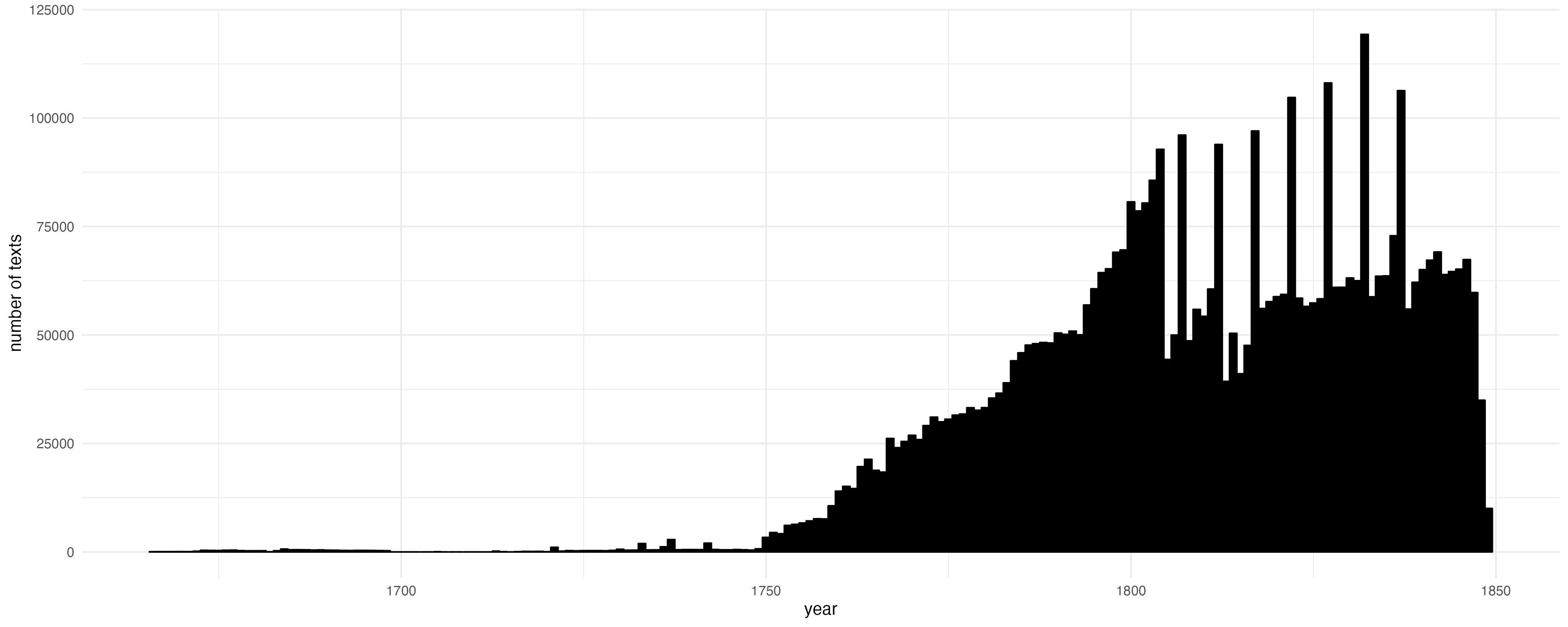}

}

\caption{Texts per year.}

\end{figure}

\hypertarget{from-pages-to-texts}{%
\subsection{From pages to texts}\label{from-pages-to-texts}}

One of our aims with the dataset was to explore certain genres of
advertisements. These include advertisements relating to specific forms
of consumption, advertisements for labour, and runaway notices. By
having a broad corpus we wanted to extract texts of relevance on a scale
that allowed for investigation of developments over time. However, one
obstacle needed to be overcome. While the images themselves already came
with metadata on a given edition, including its date of publication, we
wanted to work with standalone texts. We saw this as the natural entity
for the data given our ambitions. An edition might contain many texts,
and a text might span multiple pages. Thus, we had to devise a way to
segment the output into meaningful units.

We decided to approach this as a problem of classifying every line of
text, since lines were separated in the output of the text recognition
process. If one can infer whether a line represents the beginning of a
text, it is easy to split editions into smaller, cohesive units.
Initially, we used structural features to make this prediction,
including elements such as grammar (for instance, does the line end with
a full stop?), the number of characters for each line, capitalization,
as well as a number of complicated regular expressions matching
recurring headlines. These elements were used as features for a random
forest model that made the final prediction. This worked reasonably
well, but much better for some publications than others.

Midway through, we re-designed the process. This was possible, since by
that point we had enough text to create a new tool in the form of a
BERT-model. This model is a fill-mask model trained on 260 million words
segmented using the described approach. It was trained in collaboration
with Matias Kokholm Appel and is available on Huggingface alongside its
various fine-tunes.\footnote{DA-BERT\_Old\_News\_V1.
  https://huggingface.co/CALDISS-AAU/DA-BERT\_Old\_News\_V1.} From this
model we derived a series of setfit-models to make predictions on
whether a line was the first line of a text, a header or the last line
of a text. This was based on a manually tagged dataset of about 30,000
lines of text. Adding these features (including information on
neighbouring lines) alongside the structural features of the former
approach to a random forest model provided a more robust segmentation
that was less fickle to small changes in formatting. On line level, the
final model has an f1-score of 98.9\% on a validation dataset of about
7,000 manually tagged lines. Of course, this score relates to each line,
so the longer the text, the more chances there are of them being split
wrong. There are also still some formats that are segmented more
successfully than others. For instance, the publication Nyeste Skilderie
af Kiøbenhavn contains frequent essays and poems that the model
struggles with, as it generally wants to split them into sub-text
units.~~

In the future we want to build even more robust models for handling
these segmentation tasks. Our hope is that as our dataset grows and is
refined, we can train better embedding-models that in turn can provide
even better tools for segmentation of the corpus. Hopefully, we can
achieve a kind of positive feedback loop, as was the case when the
initial segmentation formed the basis for a better segmentation tool.~

In a similar fashion we are currently experimenting with using the
BERT-model for correcting errors in the OCR. The results are promising,
but not ready for version 1.0 of the dataset.

\hypertarget{validation}{%
\section{\texorpdfstring{\textbf{Validation}}{Validation}}\label{validation}}

In the end our text recognition model came to include about 420,000
words of text ranging across the period. It has been made available via
Transkribus. On validation data, made up of additional transcriptions
from the newspapers included in the training material, it has an error
rate of about 0.6\% at character level. However, some corners of the
dataset contain both types of print and scans different from those in
the validation data. Most importantly, we gradually included three
Norwegian papers (from Oslo, Bergen and Tronheim). These came from newer
photographs that paradoxically were a challenge to a model trained to
deal specifically with something that looks much worse. Another batch
was derived from scans of the newspaper Nyeste Skilderie after
Kiøbenhavn, which was never part of the newspaper collection at Det
Kongelige Bibliotek, but had been photographed from local library
collections and put on Internet Archive by Niels Jensen.\footnote{\url{https://archive.org/details/@uforbederlig}}
In this case, the challenge was that many pages were somewhat warped and
distorted. To get a sense of the real world precision across the corpus,
including these atypical scans, we devised a crude and lightweight
pwa-indicator of our own tailored to eighteenth- and early
nineteenth-century Danish. This was done by creating a dictionary of
words from the period. We identified recurrent tokens in existing high
quality text corpora, including selections from canonical literature of
the period, such as the writings of N.F.S. Grundtvig. These texts were
partly located via the Archive of Danish Literature corpus publicly
available at Huggingface and partly made available by the Center for
Grundtvigforskning.\footnote{\emph{Archive of Danish Literature},
  \url{https://huggingface.co/datasets/danish-foundation-models/danish-dynaword/blob/main/data/adl/adl.md}.}
Another important source was the nationwide census of 1787 available via
Denmark's National Archives which we mined for spellings of names and
places. Finally, we combed through the 10.000 most common words in an
early version of the corpus that were not in the canonical literature or
the census and then added those among them that were actually existing
words from the period. The resulting dictionary was matched with all
word tokens in each text providing a simple indication of how the text
recognition model performs. Of course this method is somewhat simple: a
word in the recognised output can be an actual word, but not the one
that is on the page. And some words are just too rare to be included in
the dictionary sources. However, in the aggregate the approach still
represents a useful way of comparing parts of the dataset to one
another.

Subsequently, this PWA-estimate is available for each text entry.
Generally, the data is of a high and consistent quality (e.g.~figure 4).
However, we also use it to document whatever inconsistencies might be
present in a given series, with a few series seeing occasional though
dramatic drops in accuracy. In our process, we used these indicators of
lacunae to go back and check -- as they should generally be explainable,
by the text being harder to read to the human eye too. Typically they
come from either larger parts of the text being in German, or from
intermittent low-contrast or blurry scans that have tripped up either
layout and text recognition, and sometimes both. Thus, for end uses
dependent on largely correct text such as language modelling, we suggest
rather aggressive filtering on this pwa-variable.

\begin{figure}

{\centering \includegraphics{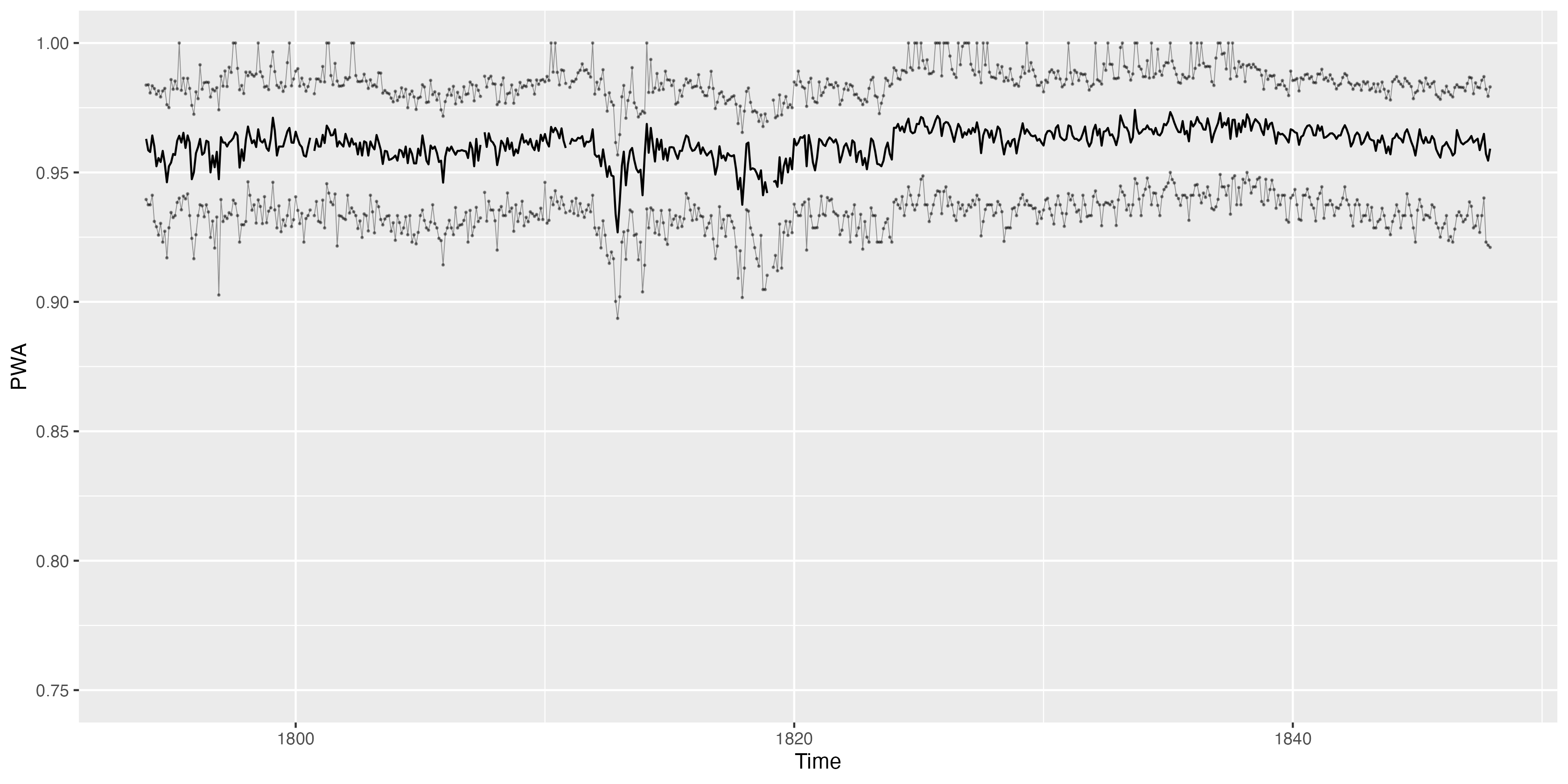}

}

\caption{Example pwa-scores. Quartiles. Aarhus Stifts-Tidende.}

\end{figure}

The relative consistency of the dataset in terms of recognition quality
shows in the distribution of pwa-scores, rendered in figure 5.

\begin{figure}

{\centering \includegraphics{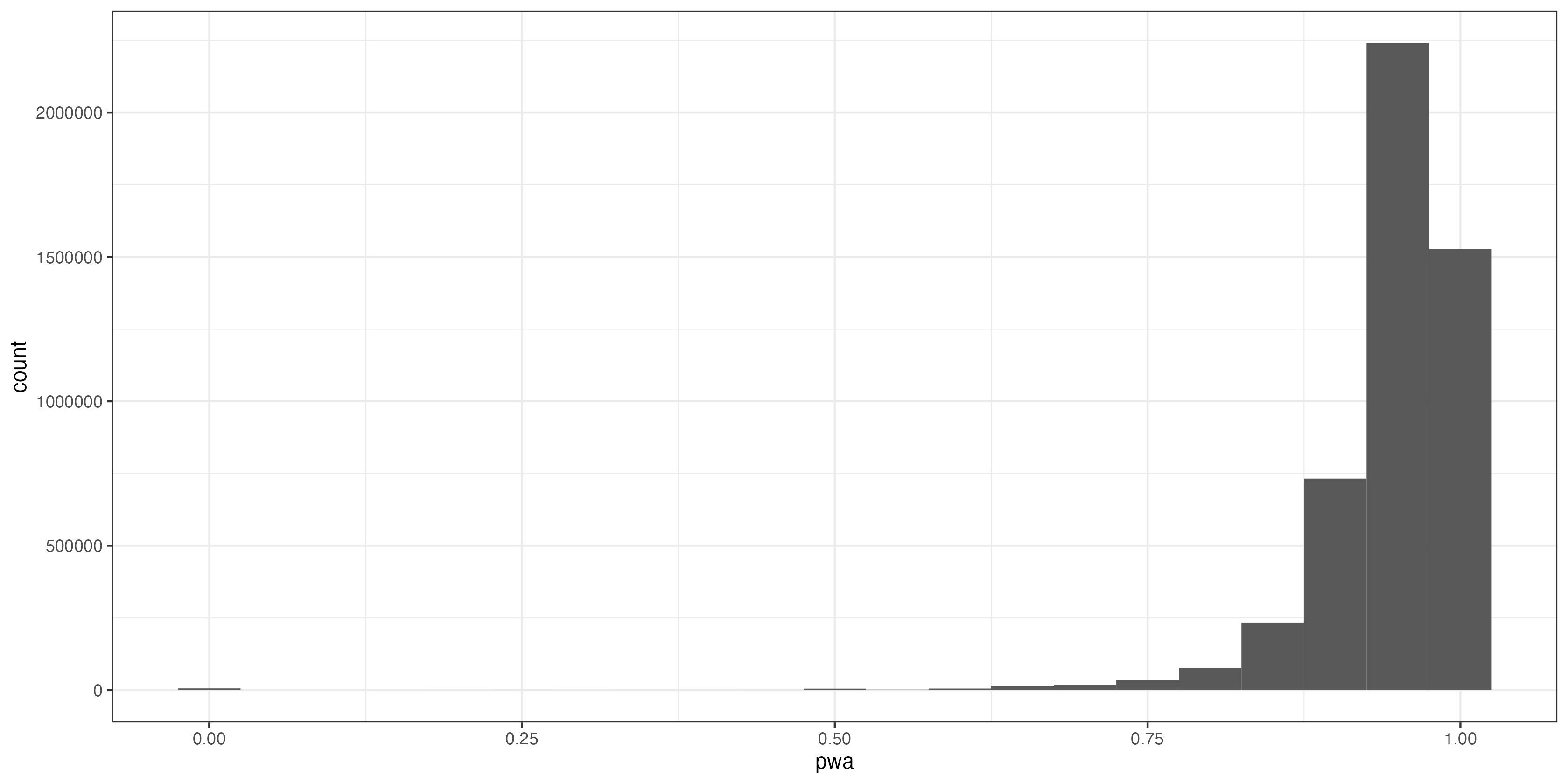}

}

\caption{Distribution of pwa-scores.}

\end{figure}

\hypertarget{usage-notes}{%
\section{\texorpdfstring{\textbf{Usage
notes}}{Usage notes}}\label{usage-notes}}

Currently, we have made the data available in two forms. We have built a
custom interface for each publication located at
\url{https://hislab.quarto.pub/}. This is meant to service both the
general public, historians, and other scientists as an alternative to
the existing library interfaces. For each publication we detail issues
with the digitisation process and provide notes on coverage. The
datasets are fully searchable and include further features. One is a
link to the relevant page in the Danish library collection, when the
data stems from there. We were informed by representatives of
Nettbiblioteket that their links are not permanent over time, and have
therefore refrained from implementing a similar trace there. However,
when available these links allow users to double check if something
seems off with the text recognition -- or if they want to look at the
digital scan of the actual print. For each text entry, we also provide
users with the ability to see similar texts in the publication. This
allows for users to find a relevant point of entry using fulltext search
or regular expression and then move horizontally through the data
following the suggestions we make. These suggestions are based on
similarity scores derived from embeddings of the texts themselves that
were made by a fine-tuned version of our BERT-model.\footnote{\emph{Old\_News\_Segmentation\_SBERT\_V0.1,}
  https://huggingface.co/JohanHeinsen/Old\_News\_Segmentation\_SBERT\_V0.1}
These embeddings are then processed using FAISS to find similar texts
based on their vector representation. For most publications, we provide
25 similar texts pr. entry, though on small publication we reduce the
number, as they might not contain enough material for there to be 25
similar elements.~

A simple text recommender like this might not sound revolutionary, but
to us it is. As trained historians we are used to our work being reigned
in by archival provenance, working within the confines of what is in one
archival box at a time. In manual archival work there is typically no
way of knowing if something similar exists elsewhere in the archive. Now
that can be tested in seconds. For newspaper archives containing
hundreds of thousands of pages on digital platforms, historians are
still required to search through an enormous amount of material.
Digitisation enabling fulltext search certainly makes this more
manageable, but this is assuming that the historians know what they are
looking for. Variations in formulations, words, and spellings, along
with a potential overwhelming number of results, poses issues for the
keyword-in-context searches.\footnote{Bingham, A. The Digitization of
  Newspaper Archives: Opportunities and Challenges for Historians.
  \emph{Twentieth Century British History} 21:2, 225-231 (2010).} This
has motivated researchers to implement different approaches such as
topic modelling on their own dataset to allow navigation between or
clustering of related texts, but having a similar function as an
integral part of an archival interface for all users takes this one step
further.

We have also made the dataset in total available via Huggingface as
either parquet or csv-files.\footnote{https://huggingface.co/datasets/JohanHeinsen/ENO.}
This version excludes links back to the original collections and the
candidates for similarity, since both added substantially to size,
making the dataset too large to fit in the RAM of your average student
laptop.

Additionally, the dataset is included in the Danish Dynaword corpus
collecting openly licensed textual data in Danish for language
modelling.\footnote{Enevoldsen, K.C., Jensen, K.N., Kostkan, J., Szab'o,
  B.I., Kardos, M., Vad, K., Heinsen, J., N'unez, A.B., Barmina, G.,
  Nielsen, J., Larsen, R., Vahlstrup, P.B., Dalum, P.M., Elliott, D.,
  Galke, L., Schneider-Kamp, P., \& Nielbo, K.L., \emph{Dynaword: From
  One-shot to Continuously Developed Datasets.} Preprint.
  https://arxiv.org/abs/2508.02271.} However, we caution that the
dataset is heavily biased in almost every conceivable way. It contains
sentiments justifying everything from beating children to selling humans
as slaves. In this capacity it should be used reflectively -- it affords
us a mirror to historical sentiments, but we should not expect those
sentiments to mirror our own.

\hypertarget{data-availability}{%
\section{\texorpdfstring{\textbf{Data
Availability}}{Data Availability}}\label{data-availability}}

The data can be accessed per publication at
\url{https://hislab.quarto.pub/eno/} and in total via
\url{https://huggingface.co/datasets/JohanHeinsen/ENO}. It is licensed
under cc-by-sa4.0.

\hypertarget{code-availability}{%
\section{\texorpdfstring{\textbf{Code
Availability}}{Code Availability}}\label{code-availability}}

All models are available on Huggingface. Additional code is available
via the project website and via the project's Github repository.

\hypertarget{acknowledgments}{%
\section{\texorpdfstring{\textbf{Acknowledgments}}{Acknowledgments}}\label{acknowledgments}}

The project received funding from MASSHINE, Center for Digital Textual
Heritage and Center for Humanities Computing. We would like to thank the
following colleagues and students for their support: Matias Kokholm
Appel, Max Odsbjerg Pedersen, Kamilla Matthiassen, Sofus Landor Dam,
Anders Dyrborg Birkemose, Bo Poulsen, Maria Simonsen, Louise Karoline
Sort, Louise Emilie Pedersen, Adrian Ledaal Gundersen, Laura Aagaard,
Rasmus Due Bak, Mads Lehrmann Christensen, Mie Dodensig, Dionysia Horne,
Maja Stecher Jensen, Ravn Jørgensen, André Sigaard Mikkelsen, Caisa
Johanna Ingerslev Nedergaard, Frederik Sørig Søndergaard Nielsen,
Christian Friis Olesen, Fiona Gesine Otten, Anne Katrine Holm Pedersen,
Maja Sjørslev Petersen, Louise Haldrup Riske, Zenia Skytte Sørensen,
Maja Kromann Sørensen, Jeppe Risager Ubbesen, Laura Marie Ahrensbach,
Julie Edelsten, Alexander Simon Kjølby Carlsen, Andreas Winkler
Bønnelykke, Jonathan Eskerod Qvistorff Kanstrup, Sarah Lydia Blok
Kloster, Mads Meldgaard Skibsted Kristensen, Magnus Østergaard Larsen,
Hans Niklas Holmgaard Pedersen, Lea Ruess, Maja Thorsø Rønn, Ditte
Nørgaard Schrøder, Oliver Thomsen, Stinna Victoria Østergaard.

\hfill\break
\hfill\break
\hfill\break
\hfill\break

\end{document}